\documentclass{JHEP3}
\usepackage{amsmath}
\usepackage{epsfig}
\usepackage{amssymb}
\usepackage{graphics}
\usepackage[active]{srcltx}

\newcommand{\bea}{\begin{eqnarray}}
\newcommand{\eea}{\end{eqnarray}}
\newcommand{\bean}{\begin{eqnarray*}}
\newcommand{\eean}{\end{eqnarray*}}
\newcommand{\nn}{\nonumber\\}

\def\W#1{\widetilde{#1}}
\def\WH#1{\widehat{#1}}

\def\braket#1{\left\langle#1\right\rangle}

\def\ket#1{\left|#1\right\rangle}
\def\gb#1{\left\langle#1\right]}
\def\tgb#1{\left[#1\right\rangle}

\def\eref#1{(\ref{#1})}

\def\c{{\gamma}}

\def\la{\lambda}

\def\vev{\braket}
\def\tgb#1{\left[#1\right\rangle}
\def\bket#1{\left|#1\right]}
\def\bvev#1{\left[#1\right]}
\def\Spaa{\vev}
\def\Spbb{\bvev}
\def\Spab{\gb}
\def\Spba{\tgb}
\def\und{\underline}

\def\Label#1{\label{#1}
\smash{\hbox to0pt{\raise1ex\hbox{\tiny[#1]}\hss}}}

\title{Determination of Boundary Contributions in Recursion Relation}

\author{Bo Feng$^{ab}$, Kang Zhou$^{a}$, Chenkai Qiao$^{a}$,
Junjie Rao$^{a}$\footnote{Corresponding author: raojunjie@zju.edu.cn}~\footnote{The
unconventional order of authors is merely to satisfy the outdated requirement
for Phy. Degree of the school.}\bigskip \\
{$^a$\small Zhejiang Institute of Modern Physics, Zhejiang University, Hangzhou, 310027, P. R. China \\
$^b$\small Center of Mathematical Science, Zhejiang University, Hangzhou, 310027, P. R. China \\}}

\abstract{In this paper, we propose a new algorithm to systematically
determine the missing boundary contributions,
when one uses the BCFW on-shell recursion relation to calculate tree amplitudes
for general quantum field theories. After an instruction of the algorithm,
we will use several examples to demonstrate its application, including amplitudes of color-ordered
$\phi^4$ theory, Yang-Mills theory, Einstein-Maxwell theory and color-ordered Yukawa theory
with $\phi^4$ interaction.}


\begin{document}

\section{Introduction}

Inspired by Witten's twistor program \cite{Witten:2003nn},
a powerful approach to calculate tree amplitudes is developed in
\cite{Britto:2004ap, Britto:2005fq}\footnote{For more details,
see reviews \cite{Bern:2007dw, Feng:2011np, Elvang:2013cua} and
corresponding citations.}. When applying this newly discovered
on-shell recursion relation, the large $z$ behavior of amplitudes
under a deformation parameterized by $z$ is crucial. For an
amplitude $A$, if $\lim_{z\to\infty}A(z)=0$, it can be nicely
reconstructed by sewing lower-point on-shell amplitudes. However, if
$\lim_{z\to\infty}A(z)\neq0$, nontrivial
boundary contributions arise which, in general, cannot be reconstructed
recursively. The analysis of large $z$ behavior is a nontrivial
issue since naive power counting of $z$ based on Feynman diagrams
may lead to wrong conclusions in many cases. A nice way to tackle this by
applying the background field method is presented in
\cite{ArkaniHamed:2008yf} by Arkani-Hamed and Kaplan. In this
way, it has been shown \cite{ArkaniHamed:2008yf, Cheung:2008dn}
that when the amplitude contains at least one gluon  or graviton, there
is at least one deformation with convergent (which means good) large $z$
behavior\footnote{More accurately, one needs other conditions, such as
two derivative theories and spins of other particles should be less than one
for the case of gluon, and two for the case of graviton.}.
However, for theories involving only scalars and fermions, or
some effective theories, boundary contributions are
unavoidable. For example, the Yukawa theory, is part of the Standard Model. Thus it is necessary
to generalize the on-shell recursion relation to deal with these terms.

Several proposals have been made to handle this difficult task. The
first \cite{Benincasa:2007xk,Boels:2010mj} is to introduce auxiliary
fields so that in the enlarged theory, there are no boundary
contributions. After working out the parent amplitudes, by proper
reduction one gets the desired derivative amplitudes. But there are two
problems: Firstly it is unknown in general whether
the enlarged theory exists, or how to construct it if it exists.
Secondly the parent amplitudes could be far more complicated than
expected, thus this way is not quite efficient. The second
\cite{Feng:2009ei,Feng:2010ku,Feng:2011twa} is to carefully analyze
Feynman diagrams and then isolate their boundary contributions,
which can be evaluated directly or recursively afterwards. However, this
approach is useful only when boundary contributions are located on
merely a few Feynman diagrams. The third \cite{Benincasa:2011kn,
Benincasa:2011pg, Feng:2011jxa} is to express boundary contributions
in terms of roots of amplitudes, which is a fascinating idea, but to
find roots is an extremely challenging work.

In this paper, we introduce a systematic algorithm to determine boundary
contributions for general quantum field theories. The key point is simple: Similar to
tree amplitudes, the boundary contributions are also rational
functions of external momenta. Thus, after carefully analyzing their
pole structure, one can capture these elusory quantities by applying exactly the same idea
used to derive the on-shell recursion relation.
At this point, it is important to restrict the algorithm with certain conditions.
Since it is based on pole structure, boundary contributions that do not
contain any poles cannot be determined.
In general the mass dimension of $n$-point amplitudes is $(4-n)$,
thus for theories without coupling constants of negative mass dimension,
there is at least one pole for $n>4$. But for some effective theories,
it is imaginable that such counterexamples could appear and then
one needs alternative approaches.

The paper is organized as follows. In section 2, we present
the general framework of this algorithm. In section 3, we use
several examples to demonstrate its versatility in applications. In
section 4, we give a brief summary and several further directions.
In appendix A, the new algorithm is reinterpreted in a more concise algebraic language.

\section{The New Algorithm}

In this section, we will present the new algorithm in detail.
Assume that all external legs and internal propagators are massless,
then one can use the spinor formalism to simplify the calculation,
although this requirement is not essential.
From now on, let's adopt the QCD convention, \textit{i.e.}, $s_{ij}=\Spaa{i|j}\Spbb{j|i}$.
Furthermore, the $i$'s in usual definitions of amplitude and propagator are dropped,
\textit{i.e.}, ${i\over P^2}\to{1\over P^2}$.
Thus the familiar BCFW recursion relation reads as $-A\sim\sum{A_LA_R\over P^2}$.

Now let's recall the on-shell recursion relation, starting by deforming a pair of spinors, \textit{e.g.},
$\und{0}\equiv\Spab{i_0|j_0}$ with $\ket{i_0}\to\ket{i_0}-z_{\und{0}}\ket{j_0}$
and $\bket{j_0}\to\bket{j_0}+z_{\und{0}}\bket{i_0}$\footnote{Since multiple deformations will be used,
to simplify symbols we adopt the following notations: $\und{s}=\Spab{i_s|j_s}$
represents the $s$-th deformation and its parameter $z_{\und{s}}$.
The location of its pole associated with $P^2$ reads as $z_{\und{s},P^2}$.}.
Under deformation $\und{0}$, all physical propagator\footnote{More strictly, physical poles
are defined to have non-vanishing factorization limits. And all other poles are spurious.}
$P_i^2$'s are divided into two categories: the {\bf detectable propagators} which depend on $z_{\und{0}}$,
and the {\bf undetectable propagators} which do not. These two sets are
denoted as ${\cal D}^{\und{0}}$ and ${\cal U}^{\und{0}}$ respectively, where the superscript indicates
the deformation. There will be also spurious poles and their set is denoted as ${\cal S}^{\und{0}}$.
As a rational function of $z_{\und{0}}$, the amplitude obtained by Feynman rules can be decomposed as
\bea
-A^{\und{0}}(z_{\und{0}})
={N(z_{\und{0}})\over\prod P_t^2(z_{\und{0}})}=\sum_{P_t^2\in{\cal D}^{\und{0}}}
{A_{t;L}(\WH z_{t,\und{0}})A_{t;R}(\WH z_{t,\und{0}})\over P_t^2(z_{\und{0}})}
+C_0^{\und{0}}+\sum C_i^{\und{0}}z^i_{\und{0}}.~~~~\label{Az0-gen}
\eea
For later convenience, we define the
{\bf recursive part}\footnote{Since for the recursive part, $z_{\und{0}}$ appears in all poles, so
it is called the `pole part' as well.} as
\bea
{\cal R}^{\und{0}}(z_{\und{0}})=\sum_{P_t^2\in{\cal D}^{\und{0}}}{A_{t;L}(\WH z_{\und{0},t})
A_{t;R}(\WH z_{\und{0},t})\over P_t^2(z_{\und{0}})},~~~~\label{Az0-gen-Rpart}
\eea
and the {\bf boundary part}\footnote{The boundary part is called the `regular part' as well.} as
\bea
{\cal B}^{\und{0}}(z_{\und{0}})
=C_0^{\und{0}}+\sum C_i^{\und{0}}z^i_{\und{0}}~.~~~\label{Az0-gen-Bpart}
\eea
Setting $z_{\und{0}}=0$ to conceal the deformation after decomposition,
let's simply denote the 0th recursive and boundary parts as
${\cal R}^{\und{0}}$ and ${\cal B}^{\und{0}}$ respectively.
It is obvious that the numerator of ${\cal R}^{\und{0}}$ is
a product of left and right on-shell amplitudes evaluated
at $P_t^2(z_{\und{0}})=0$, which can be calculated recursively, while
${\cal B}^{\und{0}}$ is nothing but the boundary contribution we seek for.
Detailed analysis on expression \eref{Az0-gen} informs us the following important facts:
\begin{itemize}

\item (A-1) Coefficients $C_0^{\und{0}}$, $C_i^{\und{0}}$ and $A_{t;L}
    (\WH z_{\und{0},t})A_{t;R}(\WH z_{\und{0},t})$ are all simply rational functions of
    spinors $\la_i$,$\W\la_i$. To understand these coefficients,
    it is crucial to determine their pole structure.

\item (A-2) It is well known that term by term, there may be {\bf spurious poles}
    in $A_{t;L}(\WH z_{\und{0},t})A_{t;R}(\WH z_{\und{0},t})$. Some spurious poles will cancel each
    other when we sum over $t$, but others will still remain.
    To cancel them for a physical amplitude, ${\cal B}^{\und{0}}$ must also depend on
    the same spurious poles. But if all of them are canceled in ${\cal R}^{\und{0}}$,
    ${\cal B}^{\und{0}}$ does not need to have this dependence any more.

\item (A-3) Now a key observation is that by such formulation, physical poles $P_t^2\in{\cal D}^{\und{0}}$
    will appear once and only once with power one in ${\cal R}^{\und{0}}$. In other words,
    {\sl they cannot be the poles of boundary contribution ${\cal B}^{\und{0}}$
    and sub-amplitudes $A_{t;L}(\WH z_{\und{0},t})A_{t;R}(\WH z_{\und{0},t})$}.

\item (A-4) Having excluded $P_t^2\in{\cal D}^{\und{0}}$ from ${\cal B}^{\und{0}}$,
    now we have a clear picture of its pole structure: (1) It must be either a
    physical or spurious pole which belongs to ${\cal U}^{\und{0}}$ or ${\cal S}^{\und{0}}$;
    (2) The powers of spurious poles in ${\cal B}^{\und{0}}$ may be larger than one.
    In fact, their degrees are determined by the corresponding degrees in
    coefficients $A_{t;L}(\WH z_{\und{0},t})A_{t;R}(\WH z_{\und{0},t})$.

\end{itemize}
Understanding the pole structure of
${\cal B}^{\und{0}}$, it is natural to determine it by using
other deformations. To proceed, let's perform a new
deformation $\und{1}\equiv\Spab{i_1|j_1}$. The only condition for the new deformation
is\footnote{While keeping ${\cal D}^{\und{0}}$ to denote only physical poles
detected by deformation $\Spab{i_0|j_0}$, we now use ${\cal D}^{\und{01}}$ to denote
poles that have been detected by two consequent deformations (thus including
spurious poles in ${\cal S}^{\und{0}}$ detected by deformation $\Spab{i_1|j_1}$).}
that it can detect at least one pole in $({\cal U}^{\und{0}}\bigcup{\cal S}^{\und{0}})$
(which means one pole in ${\cal B}^{\und{0}}$).
Obviously, in practice it is better to choose a new deformation which maximizes
the number of detected poles in $({\cal U}^{\und{0}}\bigcup{\cal S}^{\und{0}})$.

Under deformation $\und{1}$, one can write the full amplitude as a function of $z_{\und{1}}$ by two
different ways. The first is to use the expression given by
Feynman rules directly as \eref{Az0-gen}, namely
\bea
-A^{\und{1}}(z_{\und{1}})
={N(z_{\und{1}})\over\prod P_r^2(z_{\und{1}})}=\sum_{P_r\in{\cal D}^{\und{1}}}{A_{r;L}(\WH z_{\und{1},r})
A_{r;R}(\WH z_{\und{1},r})\over P_r^2(z_{\und{1}})}
+C_0^{\und{1}}+\sum C_i^{\und{1}}z^i_{\und{1}}={\cal R}^{\und{1}}(z_{\und{1}})
+{\cal B}^{\und{1}}(z_{\und{1}}).~~~~\label{Az1-gen}
\eea
The second is to use \eref{Az0-gen}, \eref{Az0-gen-Rpart} and
\eref{Az0-gen-Bpart} to perform deformation $\und{1}$, then
\bea
-A^{\und{1}}(z_{\und{1}})={\cal R}^{\und{0}}(z_{\und{1}})
+{\cal B}^{\und{0}}(z_{\und{1}}).~~~~\label{Az0-gen-defor}
\eea
Obviously, as a rational function of $z_{\und{1}}$, expression \eref{Az1-gen}
must equal to \eref{Az0-gen-defor}. Although there are unknown terms in both
\eref{Az1-gen} and \eref{Az0-gen-defor}, their recursive parts ${\cal R}^{\und{0}}(z_{\und{1}})$
and ${\cal R}^{\und{1}} (z_{\und{1}})$ are known, thus we can determine part of
${\cal B}^{\und{0}}(z_{\und{1}})$ as the following. Since
${\cal B}^{\und{0}}(z_{\und{1}})$ is a rational function of $z_{\und{1}}$,
it can be decomposed into the recursive part (\textit{i.e.}, the pole part)
and boundary part as\footnote{The order in symbol
${\cal BR}^{\und{0},\und{1}}$ makes a difference. It means the 1st recursive part of the 0th boundary part.
In contrast, ${\cal RB}^{\und{0},\und{1}}$ means the 1st boundary part of the 0th recursive part.}
\bea
{\cal B}^{\und{0}}(z_{\und{1}})={\cal BR}^{\und{0},\und{1}}(z_{\und{1}})
+{\cal B}^{\und{0}\und{1}}(z_{\und{1}}),~~~\label{C0-exp}
\eea
where
\bea
{\cal BR}^{\und{0},\und{1}}(z_{\und{1}})&=&\sum_{P_{t}^2\in
{\cal U}^{\und{0}}\bigcap{\cal D}^{\und{0}\und{1}}}\sum_{a=1}^{n_{P_t^2}}{c_{t,a}\over(P_{t}^2(z_{\und{1}}))^a}
+\sum_{S_{t}\in{\cal S}^{\und{0}}\bigcap{\cal D}^{\und{0}\und{1}}}
\sum_{b=1}^{n_{S_t}}{d_{t,b}\over(S_{t}(z_{\und{1}}))^b},\nn
{\cal B}^{\und{0}\und{1}}(z_{\und{1}})&=&{\cal B}^{\und{0}\und{1}}
+\sum C_{0i}^{\und{0}\und{1}}z^i_{\und{1}},~~~\label{C0-pole}
\eea
where $a,b$ are the degrees of corresponding poles.
In decomposition \eref{C0-exp}, note that ${\cal B}^{\und{0}\und{1}}$
no longer contains poles in $({\cal U}^{\und{0}}\bigcup{\cal S}^{\und{0}})\bigcap{\cal D}^{\und{01}}$,
although it may produce new spurious poles. We now use ${\cal U}^{\und{01}}$ to denote
the remaining undetectable physical poles, and ${\cal S}^{\und{01}}$ to denote the remaining
undetectable spurious poles in ${\cal S}^{\und{0}}$ as well as newly generated spurious poles.

Similarly, one can decompose ${\cal R}^{\und{0}}(z_{\und{1}})$ in
\eref{Az0-gen-defor} into the recursive part and boundary part as
\bea
{\cal R}^{\und{0}}(z_{\und{1}})={\cal RR}^{\und{0},\und{1}}(z_{\und{1}})
+{\cal RB}^{\und{0},\und{1}}(z_{\und{1}}),~~~~\label{AA-pole}
\eea
where both ${\cal RR}^{\und{0},\und{1}}(z_{\und{1}})$ and
${\cal RB}^{\und{0},\und{1}}(z_{\und{1}})$ are known. Now compare the recursive parts of
\eref{Az1-gen} and \eref{Az0-gen-defor}, we trivially reach an identity
\bea
{\cal R}^{\und{1}}(z_{\und{1}})={\cal RR}^{\und{0},\und{1}}(z_{\und{1}})
+{\cal BR}^{\und{0},\und{1}}(z_{\und{1}}).~~~\label{Pole-comp}
\eea
From this one can determine coefficients $c_{t,a}$ and $d_{t,b}$.
Explicitly, we have the following observations:
\begin{itemize}

\item (B-1) Firstly, since spurious poles $S_{t}(z_{\und{1}})$ do not appear in
    ${\cal R}^{\und{1}}(z_{\und{1}})$, they must be canceled by the identical terms in
    ${\cal RR}^{\und{0},\und{1}}(z_{\und{1}})$ and ${\cal BR}^{\und{0},\und{1}}(z_{\und{1}})$
    so that we can determine coefficients $d_{t,b}$ in ${\cal BR}^{\und{0},\und{1}}(z_{\und{1}})$
    from the known ${\cal RR}^{\und{0},\und{1}}(z_{\und{1}})$.
    Similarly we can determine coefficients $c_{t,a}$ by subtracting corresponding
    contributions by ${\cal RR}^{\und{0},\und{1}}(z_{\und{1}})$
    for poles $P_{t}^2\in{\cal U}^{\und{0}}\bigcap{\cal D}^{\und{1}}$ in ${\cal R}^{\und{1}}(z_{\und{1}})$.

\item (B-2) Secondly, for physical poles $P_{t}^2\in{\cal D}^{\und{0}}\bigcap{\cal D}^{\und{1}}$,
    since ${\cal BR}^{\und{0},\und{1}}(z_{\und{1}})$ does not contribute,
    contribution from ${\cal R}^{\und{1}}(z_{\und{1}})$ must equal to the
    one from ${\cal RR}^{\und{0},\und{1}}(z_{\und{1}})$.
    This serves as an important consistency check of the algorithm.

\end{itemize}
In practice, there is no need to determine coefficients $c_{t,a},d_{t,b}$ separately,
instead one can directly write
\bea
{\cal BR}^{\und{0},\und{1}}(z_{\und{1}}=0)={\cal R}^{\und{1}}(z_{\und{1}}=0)
-{\cal RR}^{\und{0},\und{1}}(z_{\und{1}}=0).~~~~~\label{C-pole}
\eea
Plugging this back, the full amplitude \eref{Az0-gen} becomes
\bea
-A={\cal R}^{\und{01}}+{\cal B}^{\und{01}},~~~~~~~
{\cal R}^{\und{01}}={\cal R}^{\und{0}}(z_{\und{0}}=0)
+{\cal R}^{\und{1}}(z_{\und{1}}=0)-{\cal RR}^{\und{0},\und{1}}(z_{\und{1}}=0),~~~~\label{Az0-gen-1}
\eea
where the unknown ${\cal B}^{\und{01}}$ can only contain poles in
$({\cal U}^{\und{01}}\bigcup{\cal S}^{\und{01}})$.
And ${\cal BR}^{\und{0},\und{1}}$ can be further simplified,
as will be explained in the end of this section.

What do we have achieved after performing the new deformation $\und{1}$?
Before this, we know that the unknown ${\cal B}^{\und{0}}$ can only contain poles in
${\cal U}^{\und{0}}\bigcup{\cal S}^{\und{0}}$. After this, we find the
${\cal BR}^{\und{0},\und{1}}$ part while the remaining unknown part ${\cal B}^{\und{01}}$
can only contain poles in $({\cal U}^{\und{01}}\bigcup{\cal S}^{\und{01}})$,
where ${\cal U}^{\und{01}}$ is a subset of ${\cal U}^{\und{0}}$.

Now one sees the recursive pattern: Treating ${\cal R}^{\und{01}},{\cal B}^{\und{01}}$
in \eref{Az0-gen-1} as ${\cal R}^{\und{0}},{\cal B}^{\und{0}}$ in \eref{Az0-gen-defor},
we can repeat the entire procedure again. Each time performing a new deformation, we get part
of the boundary contribution, while the remaining unknown part
contains less and less physical poles. If at each time the new
deformation can detect at least one physical pole, after finite
steps, the unknown part will contain no physical poles at all,
although it can depend on spurious poles. Then one needs to check
whether all spurious poles are canceled out without the unknown part.
If this holds, we can safely drop the unknown part.
If this fails, we need to use new deformations to detect the
uncanceled spurious poles in order to determine their corresponding terms.
Repeat same procedures until one has excluded all dependence on
spurious poles. Since all pole parts have been found, the remaining
part must be zero (up to possible polynomial terms),
then the boundary contribution is fully determined.
The use of auxiliary deformations has also appeared in the study of
one-loop rational parts in \cite{Bern:2005hs, Bern:2005cq,
Berger:2006ci}. In some sense, the auxiliary deformations bring
poles at infinity to finite locations, so that they can be
calculated recursively\footnote{We would like to thank David Kosower
for suggesting this viewpoint.}.

Before ending this section, as advertised before, we now
demonstrate how to simplify ${\cal BR}^{\und{0},\und{1}}$ in \eref{C-pole}.
The first trick is: From item (B-2), one can see that for
${\cal R}^{\und{1}}$ in \eref{C-pole} we don't need to calculate all poles in
${\cal D}^{\und{1}}$, but only $P_r^2$'s that belong to
${\cal D}^{\und{1}}$ and not ${\cal D}^{\und{0}}$. Similarly, when
to calculate ${\cal RR}^{\und{0},\und{1}}$ in \eref{C-pole},
one should neglect the pole part that belongs to
${\cal D}^{\und{0}}\bigcap{\cal D}^{\und{1}}$. For complicated
cases, this can save considerable amount of calculation.

The second trick is to note that sometimes
${\cal RR}^{\und{0},\und{1}}(z_{\und{1}})$ in \eref{AA-pole} is quite
difficult to directly extract from ${\cal R}^{\und{0}}(z_{\und{1}})$,
while to extract ${\cal RB}^{\und{0},\und{1}}(z_{\und{1}})$ is easy, thus one can use
\bea
{\cal RR}^{\und{0},\und{1}}(z_{\und{1}})={\cal R}^{\und{0}}(z_{\und{1}})
-{\cal RB}^{\und{0},\und{1}}(z_{\und{1}}).~~~\label{reg-exp}
\eea
Plugging this back into \eref{Az0-gen-1}, we obtain
\bea
-A&=&{\cal R}^{\und{01}}+{\cal B}^{\und{01}},\nn
{\cal R}^{\und{01}}&=&{\cal R}^{\und{0}}(z_{\und{0}}=0)
-({\cal R}^{\und{0}}(z_{\und{1}}=0)+{\cal RB}^{\und{0},\und{1}}(z_{\und{1}}=0))
+{\cal R}^{\und{1}}(z_{\und{1}}=0)\nn
&=&{\cal RB}^{\und{0},\und{1}}+{\cal R}^{\und{1}}.~~~~~\label{Az0-gen-2}
\eea
Between \eref{Az0-gen-1} and \eref{Az0-gen-2}, which choice is simpler depends on which of
${\cal RR}^{\und{0},\und{1}}(z_{\und{1}})$ and
${\cal RB}^{\und{0},\und{1}}(z_{\und{1}})$ is easier to calculate.

Let's give a brief remark.
Although we have assumed the calculation will stop at finite steps, we don't have a rigorous proof yet.
Also, there are many different choices of BCFW deformations, and among those
how many of them will be minimally enough to completely determine the
boundary contribution, and how to arrange their order will work most efficiently?
In this paper we will not try to answer these questions,
but a deeper systematic study is certainly needed in the future.

In appendix A, we will reinterpret the new algorithm in a more abstract but
concise language in terms of operators.

\section{Examples}

In this section, we present various applications to demonstrate the new algorithm.

\subsection{Six-point amplitude of color-ordered $\phi^4$ theory}

The first example is a simple six-point amplitude
of color-ordered $\phi^4$ theory \cite{Feng:2009ei}. For this case,
all possible physical poles are $P_{i(i+1)(i+2)}$ with $i=1,2,3$.
Let's start with bad deformation $\und{0}=\Spab{6|1}$
(\textit{i.e.}, it has nonzero boundary contribution), then the
amplitude is\footnote{All $i$'s have been removed from coupling constants for brevity.}
\bea
-A_6={\cal R}^{\und{0}}+{\cal B}^{\und{0}},~~~~~
{\cal R}^{\und{0}}={\la^2\over P_{123}^2},~~~\label{6-scalar-1}
\eea
with the corresponding sets ${\cal D}^{\und{0}}=\{P_{123}^2\}$, ${\cal U}^{\und{0}}
=\{P_{234}^2,P_{345}^2\}$, ${\cal S}^{\und{0}}=\emptyset$. Next, we perform deformation
$\und{1}=\Spab{4|6}$ since it can detect both poles in
${\cal U}^{\und{0}}$, and its recursive part is
\bea
{\cal R}^{\und{1}}(z_{\und{1}})
={\la^2\over P_{234}^2(z_{\und{1}})}+{\la^2\over P_{345}^2(z_{\und{1}})},
\eea
which is simply the LHS of \eref{Pole-comp}. For its RHS,
from ${\cal R}^{\und{0}}$ in \eref{6-scalar-1} we find ${\cal RR}^{\und{0},\und{1}}=0$, thus
\bea
{\cal BR}^{\und{0},\und{1}}(z_{\und{1}})={\cal R}^{\und{1}}(z_{\und{1}})
={\la^2\over P_{234}^2(z_{\und{1}})}+{\la^2\over P_{345}^2(z_{\und{1}})}.
\eea
Applying \eref{Az0-gen-1}, we get
\bea
-A_6={\la^2\over P_{123}^2}+{\la^2\over P_{234}^2}+{\la^2\over P_{345}^2}+{\cal B}^{\und{01}}.
\eea
In above expression, the known part satisfies three
criteria: (1) It does not contain any spurious poles. (2) It contains all physical poles.
(3) It satisfies all factorization limits. Thus we can finally set
${\cal B}^{\und{01}}=0$. In other words, we have determined
the boundary contribution ${\cal B}^{\und{0}}$.

\subsection{Five-point gluon amplitude $A_5(1^-, 2^+, 3^+, 4^-, 5^+)$}

The second example is a well known MHV gluon amplitude $A_5(1^-,2^+,3^+,4^-,5^+)$,
so it is easy to check its answer by the new algorithm. For
this case, all physical poles are $P_{i(i+1)}^2$ with $i=1,2,3,4,5$.
However, for a two-particle pole, one must factorize it into the
spinor and anti-spinor parts. By factorization limits only
holomorphic poles exist, thus all possible physical poles are
$\Spaa{i|i+1}$ with $i=1,2,3,4,5$. We start with bad deformation
$\und{0}=\Spab{1|5}$ which does have boundary contribution, then the amplitude is
\bea
-A^{\und{0}}_5(1^-,2^+,3^+,4^-,5^+)={\cal R}^{\und{0}}+{\cal B}^{\und{0}},~~~~~~~
{\cal R}^{\und{0}}={-\Spaa{5|1}^3\Spaa{4|2}^4\over\Spaa{1|2}\Spaa{2|3}
\Spaa{3|4}\Spaa{4|5}\Spaa{5|2}^4},~~~\label{gluon-5p-1}
\eea
with the corresponding sets
\bea
{\cal D}^{\und{0}}=\{\Spaa{1|2}\},~~~
{\cal U}^{\und{0}}=\{\Spaa{2|3},\Spaa{3|4},\Spaa{4|5},\Spaa{5|1}\},~~~
{\cal S}^{\und{0}}=\{\Spaa{2|5}^4\}.~~~\label{gluon-5p-1-set}
\eea
Note that the power of above spurious pole is four.

Now we try to perform a new deformation to detect as many poles as
possible in ${\cal U}^{\und{0}}$. One such choice is bad
deformation $\und{1}=\Spab{4|2}$, and its recursive part is directly calculated as
\bea
{\cal R}^{\und{1}}(z_{\und{1}})={\Spaa{1|5}^3\Spaa{4|2}^3\over\Spaa{1|2}\Spaa{2|3}\Spaa{3|5}\Spaa{5|2}^3}
{1\over\Spaa{4|5}-z_{\und{1}}\Spaa{2|5}}+{\Spaa{4|2}^3\Spaa{3|1}^4\over
\Spaa{1|2}\Spaa{3|5}\Spaa{5|1}\Spaa{3|2}^4}{1\over\Spaa{3|4}-z_{\und{1}}\Spaa{3|2}},
\eea
while the pole part of ${\cal R}^{\und{0}}$ from \eref{gluon-5p-1} under $\und{1}$ is given by
\bea
{\cal RR}^{\und{0},\und{1}}(z_{\und{1}})={\Spaa{5|1}^3\Spaa{4|2}^3\over\Spaa{1|2}\Spaa{2|3}
\Spaa{5|2}^4\Spaa{3|5}}\left({\Spaa{2|5}\over\Spaa{4|5}
-z_{\und{1}}\Spaa{2|5}}-{\Spaa{3|2}\over\Spaa{3|4}-z_{\und{1}}\Spaa{3|2}}\right).
\eea
Thus applying \eref{C-pole}, we find
\bea
{\cal BR}^{\und{0},\und{1}}(z_{\und{1}})
={\Spaa{4|2}^3\over\Spaa{1|2}\Spaa{3|5}}\left({\Spaa{3|1}^4\over
\Spaa{5|1}\Spaa{3|2}^4}-{\Spaa{5|1}^3\over
\Spaa{5|2}^4}\right){1\over\Spaa{3|4}-z_{\und{1}}\Spaa{3|2}}.
\eea
Note that although $\Spaa{1|2}$ appears in the denominator of
${\cal BR}^{\und{0},\und{1}}$, it is actually excluded by the vanishing
factor ${\Spaa{3|1}^4\over\Spaa{5|1}\Spaa{3|2}^4}
-{\Spaa{5|1}^3\over\Spaa{5|2}^4}$ when $\Spaa{1|2}$ tends to zero in
the collinear limit, which satisfies the claim that
${\cal B}^{\und{0}}$ cannot contain any poles in ${\cal D}^{\und{0}}$.
Plugging this back, \eref{gluon-5p-1} becomes
\bea
&&-A^{\und{0}}_5(1^-,2^+,3^+,4^-,5^+)={\cal R}^{\und{01}}+{\cal B}^{\und{01}},\nn
&&{\cal R}^{\und{01}}={-\Spaa{5|1}^3\Spaa{4|2}^4\over\Spaa{1|2}\Spaa{2|3}
\Spaa{3|4}\Spaa{4|5}\Spaa{5|2}^4}+{\Spaa{4|2}^3\over
\Spaa{1|2}\Spaa{3|5}}\left({\Spaa{3|1}^4\over\Spaa{5|1}\Spaa{3|2}^4}-{\Spaa{5|1}^3\over
\Spaa{5|2}^4}\right){1\over\Spaa{3|4}},~~~~~\label{gluon-5p-2}
\eea
where the unknown ${\cal B}^{\und{01}}$ can only contain poles in
\bea
{\cal U}^{\und{01}}={\cal U}^{\und{0}}\bigcap{\cal U}^{\und{1}}=\{\Spaa{2|3},\Spaa{5|1}\},~~~
{\cal S}^{\und{01}}={\cal S}^{\und{0}}\bigcup{\cal S}^{\und{1}}
=\{\Spaa{2|5}^4,\Spaa{2|3}^4,\Spaa{3|5}\}.~~~~\label{gluon-5p-1-set-2}
\eea
Next, let's perform another bad deformation $\und{2}=\Spab{1|2}$, and its recursive part is
\bea
{\cal R}^{\und{2}}(z_{\und{2}})={-\Spaa{4|5}^3\Spaa{1|2}^3\over\Spaa{2|3}\Spaa{3|4}\Spaa{5|2}^4}
{1\over\Spaa{5|1}-z_{\und{2}}\Spaa{5|2}},
\eea
while the pole part of ${\cal R}^{\und{01}}$ from \eref{gluon-5p-2} gives
\bea
{\cal RR}^{\und{01,2}}(z_{\und{2}})
={\Spaa{4|2}^3\over\Spaa{3|5}\Spaa{3|4}}{\Spaa{5|3}^4\Spaa{1|2}^3\over
\Spaa{5|2}^4\Spaa{3|2}^4}{1\over\Spaa{5|1}-z_{\und{2}}\Spaa{5|2}}.
\eea
Then we get
\bea
{\cal BR}^{\und{01,2}}(z_{\und{2}})={\Spaa{1|2}^3\over\Spaa{2|3}\Spaa{3|4}\Spaa{5|2}^4}
\left({-\Spaa{4|5}^3}-{\Spaa{4|2}^3\Spaa{3|5}^3\over\Spaa{2|3}^3}\right)
{1\over\Spaa{5|1}-z_{\und{2}}\Spaa{5|2}}.
\eea
Again it is easy to see that although the denominator contains
$\Spaa{3|4}$, it is excluded by the vanishing factor
${-\Spaa{4|5}^3}-{\Spaa{4|2}^3\Spaa{3|5}^3\over\Spaa{2|3}^3}$,
which again satisfies the claim that only poles in
\eref{gluon-5p-1-set-2} can appear. Plugging this back, \eref{gluon-5p-1} becomes
\bea
&&-A^{\und{0}}_5(1^-,2^+,3^+,4^-,5^+)={\cal R}^{\und{012}}+{\cal B}^{\und{012}},\nn
&&{\cal R}^{\und{012}}={-\Spaa{5|1}^3\Spaa{4|2}^4\over\Spaa{1|2}\Spaa{2|3}
\Spaa{3|4}\Spaa{4|5}\Spaa{5|2}^4}+{\Spaa{4|2}^3\over\Spaa{1|2}\Spaa{3|5}\Spaa{3|4}}
\left({\Spaa{3|1}^4\over\Spaa{5|1}\Spaa{3|2}^4}-{\Spaa{5|1}^3\over\Spaa{5|2}^4}\right)\nn
&&~~~~~~~~
+{\Spaa{1|2}^3\over\Spaa{2|3}\Spaa{3|4}\Spaa{5|2}^4\Spaa{5|1}}
\left({-\Spaa{4|5}^3}-{\Spaa{4|2}^3\Spaa{3|5}^3\over\Spaa{2|3}^3}\right),~~~~~\label{gluon-5p-3}
\eea
where the unknown ${\cal B}^{\und{012}}$ can only contain poles in
\bea
{\cal U}^{\und{012}}=\{\Spaa{2|3}\},~~~
{\cal S}^{\und{012}}=\{\Spaa{2|5}^4,\Spaa{2|3}^4,\Spaa{3|5}\}.~~~~\label{gluon-5p-1-set-3}
\eea
Finally, there is only one physical pole $\Spaa{2|3}$ left, and we
need to choose a deformation to detect it then produce a simplest
term from \eref{gluon-5p-1-set-3}. One choice is
$\und{3}=\Spab{3|4}$ (there is no bad deformation for the last pole
$\Spaa{2|3}$), then its recursive part is
\bea
{\cal R}^{\und{3}}(z_{\und{3}})
={-\Spaa{1|4}^4\over\Spaa{1|2}\Spaa{3|4}\Spaa{4|5}\Spaa{5|1}}
{1\over\Spaa{2|3}-z_{\und{3}}\Spaa{3|4}},
\eea
while the pole part of ${\cal R}^{\und{012}}$ in \eref{gluon-5p-3} is simply
\bea
&&\left\{{-\Spaa{5|1}^3\Spaa{4|2}^4\over\Spaa{1|2}\Spaa{2|3}
\Spaa{3|4}\Spaa{4|5}\Spaa{5|2}^4}+{\Spaa{4|2}^3\over\Spaa{1|2}\Spaa{3|5}\Spaa{3|4}}
\left({\Spaa{3|1}^4\over\Spaa{5|1}\Spaa{3|2}^4}-{\Spaa{5|1}^3\over\Spaa{5|2}^4}\right)\right.\nn
&&\left.+{\Spaa{1|2}^3\over\Spaa{2|3}\Spaa{3|4}\Spaa{5|2}^4\Spaa{5|1}}
\left({-\Spaa{4|5}^3}-{\Spaa{4|2}^3\Spaa{3|5}^3\over\Spaa{2|3}^3}\right)
\right\}_{\ket{3}\to\ket{3}-z_{\und{3}}\ket{4}}.
\eea
From above one finds
\bea
&&{\cal BR}^{\und{012},\und{3}}={-\Spaa{1|4}^4\over\Spaa{1|2}
\Spaa{3|4}\Spaa{4|5}\Spaa{5|1}}{1\over\Spaa{2|3}}-\left\{{-\Spaa{5|1}^3
\Spaa{4|2}^4\over\Spaa{1|2}\Spaa{2|3}\Spaa{3|4}\Spaa{4|5}\Spaa{5|2}^4}\right.\nn
&&\left.+{\Spaa{4|2}^3\over\Spaa{1|2}\Spaa{3|5}\Spaa{3|4}}
\left({\Spaa{3|1}^4\over\Spaa{5|1}\Spaa{3|2}^4}-{\Spaa{5|1}^3\over
\Spaa{5|2}^4}\right)+{\Spaa{1|2}^3\over\Spaa{2|3}\Spaa{3|4}\Spaa{5|2}^4\Spaa{5|1}}
\left({-\Spaa{4|5}^3}-{\Spaa{4|2}^3\Spaa{3|5}^3\over\Spaa{2|3}^3}\right)\right\}.
\eea
Plugging this back, \eref{gluon-5p-1} becomes
\bea
&&-A^{\und{0}}_5(1^-,2^+,3^+,4^-,5^+)={-\Spaa{1|4}^4\over\Spaa{1|2}
\Spaa{2|3}\Spaa{3|4}\Spaa{4|5}\Spaa{5|1}}+{\cal B}^{\und{0123}},~~~~~\label{gluon-5p-4}
\eea
where the unknown ${\cal B}^{\und{0123}}$ can only contain poles in (no poles, actually)
\bea
{\cal U}^{\und{0123}}=\emptyset,~~~{\cal S}^{\und{0123}}=\emptyset.~~~~\label{gluon-5p-1-set-4}
\eea
Since there is no spurious pole in the known part of
\eref{gluon-5p-4}, we can conclude that ${\cal B}^{\und{0123}}=0$,
and this gives the correct answer.

Let's give a summary of this example. Starting from bad deformation
$\Spab{1|5}$, we choose another three deformations
$\Spab{4|2},\Spab{1|2},\Spab{3|4}$ to determine the unknown boundary
contributions. Among these, the first two are intentionally chosen
since they are both bad deformations. After each step, the number of
physical poles in ${\cal U}$ on which remaining boundary
contributions can depend, is reduced. We are forced to perform the
last good deformation by demanding that it can detect the last pole $\Spaa{2|3}$.

\subsection{Einstein-Maxwell theory}

In this subsection, we present the amplitudes in Einstein-Maxwell
theory, which dictates interaction between photons and gravitons, and among gravitons themselves.
This case has been studied in \cite{Benincasa:2011kn, Zhou:2014yaa}, where various
relevant expressions can be found.

The recursion relation starts with the primitive three-point amplitudes below,
\textit{i.e.}, the photon-photon-graviton amplitudes
\bea
A_3(1^{-1}_\c,2^{+1}_\c,3^{-2}_g)=\kappa{{\Spaa{1|3}}^4\over{\Spaa{1|2}}^2},~~~
A_3(1^{-1}_\c,2^{+1}_\c,3^{+2}_g)=\kappa{{\Spbb{2|3}}^4\over{\Spbb{1|2}}^2},~~~\label{3p-ppg}
\eea
and the three-graviton amplitudes
\bea
A_3(1^{-2}_g,2^{-2}_g,3^{+2}_g)=\kappa{{\Spaa{1|2}}^6\over{\Spaa{2|3}}^2{\Spaa{3|1}}^2},~~~
A_3(1^{+2}_g,2^{+2}_g,3^{-2}_g)=\kappa{{\Spbb{1|2}}^6\over{\Spbb{2|3}}^2{\Spbb{3|1}}^2}.~~~\label{3p-ggg}
\eea
These are the building blocks of all higher-point amplitudes.

\subsubsection{Four-point amplitude $A_4(1_\gamma^{-1},2_\gamma^{+1},3_\gamma^{-1},4_\gamma^{+1})$}

For this four-point amplitude\footnote{For four-point amplitudes, pole $\Spaa{i|j}$
is equivalent to $\Spbb{k|l}$ ($i\neq j\neq k\neq l$) by momentum conservation.},
analysis on factorization limits shows that for its helicity configuration
$(1_\gamma^{-1},2_\gamma^{+1},3_\gamma^{-1},4_\gamma^{+1})$, all possible physical poles are
$\{\Spaa{1|2},\Spbb{1|2},\Spaa{1|4},\Spbb{1|4}\}$. Let's start with
deformation $\und{0}=\Spab{2|1}$, then its recursive part is
\bea
{\cal R}^{\und{0}}=\kappa^2{\Spaa{1|4}\Spbb{2|4}^4\over\Spbb{1|4}\Spbb{2|3}^2},
\eea
with the corresponding sets ${\cal D}^{\und{0}}=\{\Spbb{1|4}\}$,
${\cal U}^{\und{0}}=\{\Spaa{1|2},\Spbb{1|2},\Spaa{1|4}\}$, ${\cal S}^{\und{0}}=\{\Spbb{2|3}^2\}$.

Next, we perform deformation $\und{1}=\Spab{4|1}$ to detect
$\Spbb{1|2}$. Its recursive part can be obtained by exchanging labels 2 and 4
of ${\cal R}^{\und{0}}$, then
\bea
{\cal R}^{\und{1}}=\kappa^2{\Spaa{1|2}\Spbb{2|4}^4\over\Spbb{1|2}\Spbb{3|4}^2}.
\eea
On the other hand, ${\cal RR}^{\und{0},\und{1}}(z_{\und{1}})=0$ under $\und{1}$
so ${\cal RB}^{\und{0},\und{1}}={\cal R}^{\und{0}}$. According to \eref{Az0-gen-2}, we get
\bea
{\cal R}^{\und{01}}={\cal R}^{\und{0}}+{\cal R}^{\und{1}}=\kappa^2{\Spaa{1|4}\Spbb{2|4}^4\over
\Spbb{1|4}\Spbb{2|3}^2}+\kappa^2{\Spaa{1|2}\Spbb{2|4}^4\over\Spbb{1|2}\Spbb{3|4}^2}.
\eea
This is already the correct answer, as one can check its factorization
limits. But strictly speaking, we still have undetected poles in
${\cal D}^{\und{01}}=\{\Spbb{1|2},\Spbb{1|4}\}$, ${\cal U}^{\und{01}}=\{\Spaa{1|2},\Spaa{1|4}\}$,
${\cal S}^{\und{01}}=\{\Spbb{2|3}^2,\Spbb{3|4}^2\}$. Then let's perform deformation
$\und{2}=\Spab{1|2}$ to detect $\Spaa{1|4}$, and its recursive part is
\bea
{\cal R}^{\und{2}}=-\kappa^2{\Spbb{1|4}\Spaa{1|2}^4\Spaa{3|4}^4\over
\Spaa{1|4}\Spaa{2|3}^2\Spaa{2|4}^4}.
\eea
To continue, there are two equivalent ways, as given in section 2. The first is to write
\bea
{\cal RR}^{\und{01},\und{2}}(z_{\und{2}})=\kappa^2{\Spaa{2|4}\Spbb{3|4}^4\Spbb{2|1}^4\over
\Spbb{1|4}\Spbb{1|3}^5(\Spbb{3|2}+z_{\und{2}}\Spbb{3|1})},
\eea
compare this with ${\cal R}^{\und{2}}$ and use \eref{C-pole},
we find ${\cal BR}^{\und{01},\und{2}}=0$, thus ${\cal R}^{\und{012}}={\cal R}^{\und{01}}$.

The second way as by \eref{Az0-gen-2}, requires the constant term
of ${\cal R}^{\und{01}}(z_{\und{2}})$, which is
\bea
{\cal RB}^{\und{01},\und{2}}=\kappa^2{4P_{13}^2\Spbb{2|1}^2\Spbb{3|4}^2
-2P_{14}^2\Spbb{3|1}^2\Spbb{4|2}^2+P_{14}^2\Spbb{3|2}^2\Spbb{4|1}^2\over\Spbb{3|1}^4}
+\kappa^2{\Spaa{1|2}\Spbb{2|4}^4\over\Spbb{1|2}\Spbb{3|4}^2}~~,
\eea
thus we get
\bea
{\cal R}^{\und{012}}={\cal RB}^{\und{01},\und{2}}+{\cal R}^{\und{2}}
={\cal R}^{\und{01}}.
\eea
Similarly, one can perform another deformation to detect
$\Spaa{1|2}$ and then confirm that there is no further boundary contribution.
Let's simplify the result into a more compact form,
since we need it to construct higher-point amplitudes, as
\bea
A_4(1_\gamma^{-1},2_\gamma^{+1},3_\gamma^{-1},4_\gamma^{+1})=
\kappa^2{P_{13}^2\Spaa{1|3}^2\Spbb{2|4}^2\over P_{12}^2P_{14}^2}~~.~~~\label{EM-4-result-1}
\eea
It is worth mentioning that in above we have only used deformations $\Spab{2|1}$ and $\Spab{4|1}$.
Each deformation has nonzero boundary contribution, as can be checked in \eref{EM-4-result-1} directly.

\subsubsection{Four-point amplitude $A_4(1^{-1}_\c,2^{+1}_\c,3^{-2}_g,4^{+2}_g)$}

For this case, analysis on factorization limits shows that all possible physical poles are
$\{\Spaa{1|2},\Spbb{1|2},\Spbb{1|3},\Spaa{1|4}\}$.
And the good deformations are $\Spab{4|1}$ and $\Spab{1|3}$,
but to demonstrate the new algorithm, only bad deformations are used here.

Starting with deformation $\und{0}=\Spab{2|1}$, we get
\bea
-A_4=-\kappa^2{{\Spaa{1|3}}^3{\Spbb{2|4}^2}\over{\Spaa{1|4}}^2\Spbb{1|3}}+{\cal B}^{\und{0}},
\eea
with the corresponding sets ${\cal D}^{\und{0}}=\{\Spbb{1|3}\}$,
${\cal U}^{\und{0}}=\{\Spaa{1|2},\Spbb{1|2},\Spaa{1|4}\}$, ${\cal S}^{\und{0}}=\{\Spaa{1|4}^2\}$.

Next, we perform deformation $\und{1}=\Spab{3|1}$, under which
the pole part of ${\cal R}^{\und{0}}$ is zero. And the recursive part of $\und{1}$ is
\bea
{\cal R}^{\und{1}}(z_{\und{1}})=\kappa^2{{\Spaa{1|3}}^2{\Spbb{2|4}}^4\over
\Spaa{1|2}{\Spbb{2|3}}^3\left(z_{\und{1}}+\frac{\Spbb{1|2}}{\Spbb{3|2}}\right)}.
\eea
From these we get
\bea
{\cal B}{\cal R}^{\und{0},\und{1}}={\cal R}^{\und{1}}-{\cal R}{\cal R}^{\und{0},\und{1}}
=\kappa^2{{\Spaa{1|3}}^2{\Spbb{2|4}}^4\over\Spaa{1|2}{\Spbb{2|3}}^3\frac{\Spbb{1|2}}{\Spbb{3|2}}},
\eea
then the amplitude becomes
\bea
-A_4={\cal R}^{\und{0}}+{\cal B}{\cal R}^{\und{0},\und{1}}+{\cal B}^{\und{01}}
=-\kappa^2{{\Spaa{1|3}}^2{\Spaa{2|3}}^2{\Spbb{2|4}}^4\over P_{12}^2P_{13}^2P_{14}^2}+{\cal B}^{\und{01}},
\eea
with the corresponding sets ${\cal D}^{\und{01}}=\{\Spbb{1|2},\Spbb{1|3}\}$,
${\cal U}^{\und{01}}=\{\Spaa{1|2},\Spaa{1|4}\}$, ${\cal S}^{\und{01}}=\emptyset$.
Now the unknown ${\cal B}^{\und{01}}$ may depend on $\Spaa{1|2}$ and $\Spaa{1|4}$,
so one should perform further deformations to detect these two.
However, since ${\cal R}^{\und{01}}$ already satisfies all
factorization limits, we can conclude that ${\cal B}^{\und{01}}=0$, and it gives the correct answer.

To check this, one can perform deformation $\und{2}=\Spab{1|4}$ to detect $\Spaa{1|2}$,
and its recursive part is
\bea
{\cal R}^{\und{2}}(z_{\und{2}})=\kappa^2{{\Spaa{2|3}}^6{\Spaa{1|4}^4\Spbb{2|1}}\over {\Spaa{2|4}}^6{\Spaa{3|4}}^2\Spaa{2|4}\left(z_{\und{2}}-\frac{\Spaa{1|2}}{\Spaa{4|2}}\right)},
\eea
then the pole part of ${\cal B}^{\und{01}}$ is given by
\bea
{\cal B}{\cal R}^{\und{01},\und{2}}={\cal R}^{\und{2}}-{\cal R}{\cal R}^{\und{01},\und{2}}=0.
\eea
From this we get
\bea
-A_4=-\kappa^2{{\Spaa{1|3}}^2{\Spaa{2|3}}^2{\Spbb{2|4}}^4\over P_{12}^2P_{13}^2P_{14}^2}
+{\cal B}^{\und{012}},
\eea
where the unknown ${\cal B}^{\und{012}}$ does not contain any poles in
${\cal D}^{\und{012}}=\{\Spaa{1|2},\Spbb{1|2},\Spbb{1|3}\}$.
Similarly, one can perform deformation $\und{3}=\Spab{1|2}$ to detect
$\Spaa{1|4}$ and confirm that there is no further boundary contribution from ${\cal B}^{\und{012}}$.

Thus, the correct answer is
\bea
A_4(1^{-1}_\c,2^{+1}_\c,3^{-2}_g,4^{+2}_g)=\kappa^2{{\Spaa{1|3}}^2{\Spaa{2|3}}^2{\Spbb{2|4}}^4
\over P_{12}^2P_{13}^2P_{14}^2}~~.~~~\label{EM-4-result-2}
\eea
Again in above only bad deformations $\Spab{2|1}$ and $\Spab{3|1}$ have been used.

\subsubsection{Five-point amplitude $A_5(1_\gamma^{-1},2_\gamma^{+1},3_\gamma^{-1},4_\gamma^{+1},5_g^{-2})$}

For this case, analysis on factorization limits shows that all possible physical poles are\\
$\{\Spbb{1|2},\Spbb{1|4},\Spbb{1|5},\Spbb{2|3},\Spbb{2|5},\Spbb{3|4},\Spbb{3|5},\Spbb{4|5}\}$.
Starting with deformation $\und{0}=\Spab{2|1}$, we find
\bea
{\cal R}^{\und{0}}=-\kappa^3{\Spaa{1|4}\Spaa{3|5}\Spbb{3|4}\Spbb{2|4}^4\over
\Spbb{1|4}\Spbb{2|3}\Spbb{2|5}\Spbb{3|5}\Spbb{4|5}}
+\kappa^3{\Spaa{1|5}\Spaa{3|4}\Spbb{3|5}\Spbb{2|4}^5\over
\Spbb{1|5}\Spbb{2|3}\Spbb{3|4}\Spbb{4|5}\Spbb{2|5}^2},
\eea
with the corresponding sets ${\cal D}^{\und{0}}=\{\Spbb{1|4},\Spbb{1|5}\}$,
${\cal U}^{\und{0}}=\{\Spbb{1|2},\Spbb{2|3},\Spbb{2|5},\Spbb{3|4},\Spbb{3|5},\Spbb{4|5}\}$,
${\cal S}^{\und{0}}=\{\Spbb{2|5}^2\}$.

Next, we perform deformation $\und{1}=\Spab{3|2}$ to detect $\Spbb{1|2}$ and $\Spbb{2|5}$,
and its recursive part is
\bea
{\cal R}^{\und{1}}=\kappa^3{\Spaa{1|2}\Spaa{4|5}\Spbb{2|3}^4\Spbb{1|4}^5\over
\Spbb{1|2}\Spbb{1|5}\Spbb{3|4}\Spbb{3|5}\Spbb{4|5}\Spbb{1|3}^4}
-\kappa^3{\Spaa{2|5}\Spaa{1|4}\Spbb{1|3}\Spbb{2|3}^4\Spbb{4|5}^5\over
\Spbb{2|5}\Spbb{1|4}\Spbb{3|4}\Spbb{1|5}\Spbb{3|5}^6}.
\eea
On the other hand, ${\cal R}{\cal B}^{\und{0},\und{1}}$ is given by
\bea
&&{\cal R}{\cal B}^{\und{0},\und{1}}=\kappa^3{\Spab{1|4|3}\Big(\Spaa{2|5}\Spbb{2|3}^4\Spbb{4|5}^4
+\Spaa{3|5}\Spbb{4|3}\Spbb{5|3}(\Spbb{4|5}\Spbb{3|2}
-\Spbb{4|2}\Spbb{5|3})(\Spbb{4|5}^2\Spbb{3|2}^2
+\Spbb{4|2}^2\Spbb{5|3}^2)\Big)\over\Spbb{3|2}\Spbb{4|1}\Spbb{5|4}\Spbb{5|3}^6}\nn
&&+\kappa^3{\Spaa{1|5}\Big(-5\Spaa{2|4}\Spbb{3|2}^4\Spbb{5|4}^4
+\Spaa{3|4}\Spbb{4|3}\Spbb{5|3}(\Spbb{4|3}^3\Spbb{5|2}^3
+5\Spbb{3|2}^3\Spbb{5|4}^3+5\Spbb{3|2}\Spbb{5|4}\Spbb{4|2}^2\Spbb{5|3}^2)\Big)\over
\Spbb{3|2}\Spbb{5|1}\Spbb{5|4}\Spbb{5|3}^5}.~~~~~
\eea
Then we get
\bea
{\cal R}^{\und{01}}&=&{\cal R}{\cal B}^{\und{0},\und{1}}+{\cal R}^{\und{1}}\nn
&=&\kappa^3{\Spab{1|4|3}\Big(\Spaa{2|5}\Spbb{2|3}^4\Spbb{4|5}^4
+\Spaa{3|5}\Spbb{4|3}\Spbb{5|3}(\Spbb{4|5}\Spbb{3|2}
-\Spbb{4|2}\Spbb{5|3})(\Spbb{4|5}^2\Spbb{3|2}^2
+\Spbb{4|2}^2\Spbb{5|3}^2)\Big)\over\Spbb{3|2}\Spbb{4|1}\Spbb{5|4}\Spbb{5|3}^6}\nn
&&+\kappa^3{\Spaa{1|5}\Big(-5\Spaa{2|4}\Spbb{3|2}^4\Spbb{5|4}^4
+\Spaa{3|4}\Spbb{4|3}\Spbb{5|3}(\Spbb{4|3}^3\Spbb{5|2}^3
+5\Spbb{3|2}^3\Spbb{5|4}^3+5\Spbb{3|2}\Spbb{5|4}\Spbb{4|2}^2\Spbb{5|3}^2)\Big)\over
\Spbb{3|2}\Spbb{5|1}\Spbb{5|4}\Spbb{5|3}^5}\nn
&&+\kappa^3{\Spaa{1|2}\Spaa{4|5}\Spbb{2|3}^4\Spbb{1|4}^5\over
\Spbb{1|2}\Spbb{1|5}\Spbb{3|4}\Spbb{3|5}\Spbb{4|5}\Spbb{1|3}^4}
-\kappa^3{\Spaa{2|5}\Spaa{1|4}\Spbb{1|3}\Spbb{2|3}^4\Spbb{4|5}^5\over
\Spbb{2|5}\Spbb{1|4}\Spbb{3|4}\Spbb{1|5}\Spbb{3|5}^6},
\eea
with the corresponding sets ${\cal D}^{\und{01}}=\{\Spbb{1|2},\Spbb{1|4},\Spbb{1|5},\Spbb{2|5}\}$,
${\cal U}^{\und{01}}=\{\Spbb{2|3},\Spbb{3|4},\Spbb{3|5},\Spbb{4|5}\}$,\\
${\cal S}^{\und{01}}=\{\Spbb{1|3}^4,\Spbb{3|5}^5,\Spbb{3|5}^6\}$.
Next, let's perform deformation $\und{2}=\Spab{4|3}$ to detect $\Spbb{2|3}$ and $\Spbb{3|5}$,
and its recursive part is
\bea
{\cal R}^{\und{2}}=-\kappa^3{\Spaa{3|2}\Spaa{1|5}\Spbb{1|2}\Spbb{4|2}^4\over
\Spbb{3|2}\Spbb{4|1}\Spbb{4|5}\Spbb{1|5}\Spbb{2|5}}
+\kappa^3{\Spaa{3|5}\Spaa{1|2}\Spbb{1|5}\Spbb{4|2}^5\over
\Spbb{3|5}\Spbb{4|1}\Spbb{1|2}\Spbb{2|5}\Spbb{4|5}^2},
\eea
while ${\cal R}{\cal B}^{\und{01},\und{2}}$ is given by
\bea
{\cal R}{\cal B}^{\und{01},\und{2}}=\kappa^3{(\Spaa{1|3}\Spaa{2|5}\Spbb{2|1}\Spbb{5|4}
-\Spaa{1|2}\Spaa{3|5}\Spbb{4|1}\Spbb{5|2})\Spbb{2|4}^4\over
\Spbb{2|1}\Spbb{4|3}\Spbb{5|1}\Spbb{5|2}\Spbb{5|4}^2}.
\eea
Summing these, we get
\bea
{\cal R}^{\und{012}}&=&{\cal R}{\cal B}^{\und{01},\und{2}}+{\cal R}^{\und{2}}\nn
&=&\kappa^3\bigg\{{(\Spaa{1|3}\Spaa{2|5}\Spbb{2|1}\Spbb{5|4}
-\Spaa{1|2}\Spaa{3|5}\Spbb{4|1}\Spbb{5|2})\Spbb{2|4}^4\over
\Spbb{2|1}\Spbb{4|3}\Spbb{5|1}\Spbb{5|2}\Spbb{5|4}^2}-{\Spaa{3|2}\Spaa{1|5}\Spbb{1|2}\Spbb{4|2}^4
\over\Spbb{3|2}\Spbb{4|1}\Spbb{4|5}\Spbb{1|5}\Spbb{2|5}}
\nn&&+{\Spaa{3|5}\Spaa{1|2}\Spbb{1|5}\Spbb{4|2}^5\over
\Spbb{3|5}\Spbb{4|1}\Spbb{1|2}\Spbb{2|5}\Spbb{4|5}^2}\bigg\}.~~~~\label{EM-result-2}
\eea
with the corresponding sets ${\cal D}^{\und{012}}=\{\Spbb{1|2},\Spbb{1|4},
\Spbb{1|5},\Spbb{2|3},\Spbb{2|5},\Spbb{3|5}\}$,
${\cal U}^{\und{012}}=\{\Spbb{3|4},\Spbb{4|5}\}$,\\
${\cal S}^{\und{012}}=\{\Spbb{4|5}^2\}$, which requires
further detections. However, ${\cal R}^{\und{012}}$ is already
the correct answer. To verify this, one can perform deformation
$\und{3}=\Spab{1|4}$ to detect $\Spbb{3|4}$ and $\Spbb{4|5}$.
After some simplification the pole part of ${\cal R}^{\und{012}}(z_{\und{3}})$
under $\und{3}$ becomes
\bea
{\cal R}{\cal R}^{\und{012},\und{3}}(z_{\und{3}})=\kappa^3{1\over z_{\und{3}}
+{\Spbb{3|4}\over\Spbb{3|1}}}{\Spaa{3|4}\Spaa{2|5}\Spbb{4|1}^4\Spbb{3|2}^5\over
\Spbb{3|1}\Spbb{3|5}\Spbb{1|2}\Spbb{1|5}\Spbb{2|5}\Spbb{3|1}^4}
-\kappa^3{1\over z_{\und{3}}+{\Spbb{4|5}\over\Spbb{1|5}}}{\Spaa{4|5}\Spaa{3|2}\Spbb{3|1}
\Spbb{4|1}^4\Spbb{2|5}^5\over\Spbb{1|5}\Spbb{3|2}\Spbb{1|2}\Spbb{3|5}\Spbb{1|5}^6},
\eea
which is the same as the recursive part of $\und{3}$, namely
\bea
{\cal R}^{\und{3}}=\kappa^3{\Spaa{3|4}\Spaa{2|5}\Spbb{4|1}^4\Spbb{3|2}^5\over
\Spbb{3|4}\Spbb{3|5}\Spbb{1|2}\Spbb{1|5}\Spbb{2|5}\Spbb{3|1}^4}
-\kappa^3{\Spaa{4|5}\Spaa{3|2}\Spbb{3|1}\Spbb{4|1}^4\Spbb{2|5}^5\over
\Spbb{4|5}\Spbb{3|2}\Spbb{1|2}\Spbb{3|5}\Spbb{1|5}^6}.
\eea
This observation implies ${\cal R}^{\und{0123}}={\cal R}^{\und{012}}$.
One can check that \eref{EM-result-2} is equivalent
to the relevant expression in \cite{Benincasa:2011kn, Zhou:2014yaa}.
Again only bad deformations have been used here.

\subsection{Color-ordered Yukawa theory}

In this subsection,
we present the color-ordered amplitudes of fermions and scalars in Yukawa theory,
where among scalars there is also $\phi^4$ coupling. Here we focus on one type of amplitudes, namely
$A_n(1_f,2_s,\ldots,(n-1)_s, n_f)$ with only one pair of fermions $(1_f,n_f)$.
This case has been studied in \cite{Feng:2009ei}.
For checking convenience, we summarize all relevant results calculated
by Feynman diagrams up to six points as below: The three-point amplitudes are
\bea
A_3(1^+,2,3^+)=g\Spbb{1|3},~~A_3(1^-,2,3^-)=g\Spaa{1|3},~~
A_3(1^+,2,3^-)=0,~~A_3(1^+,2,3^-)=0,~~\label{Yukawa-A3}
\eea
the four-point amplitudes are
\bea
A_4(1^-,2,3,4^-)&=&0,~~~~~A_4(1^+,2,3,4^+)=0,\nn
A_4(1^-,2,3,4^+)&=&g^2{\Spaa{1|2}\Spbb{2|4}\over P_{12}^2}=-g^2{\Spaa{1|3}\over\Spaa{4|3}},\nn
A_4(1^+,2,3,4^-)&=&g^2{\Spbb{1|2}\Spaa{2|4}\over P_{12}^2}
=-g^2{\Spbb{1|3}\over\Spbb{4|3}},~~~\label{Yukawa-A4}
\eea
the five-point amplitudes are
\bea
A_5(1^-,2,3,4,5^-)&=&-g^3{\Spbb{2|4}\over\Spbb{2|1}\Spbb{5|4}}+g\la{1\over\Spbb{5|1}},\nn
A_5(1^+,2,3,4,5^+)&=&-g^3{\Spaa{2|4}\over\Spaa{2|1}\Spaa{5|4}}+g\la{1\over\Spaa{5|1}},\nn
A_5(1^+,2,3,4,5^-)&=&0,~~~~~A(1^-,2,3,4,5^+)=0,~~~\label{Yukawa-A5}
\eea
and the six-point amplitudes are
\bea
A_6(1^-,2,3,4,5,6^-)&=&0,~~~~~A_6(1^+,2,3,4,5,6^+)=0,\nn
A_6(1^-,2,3,4,5,6^+)&=&g^4{\Spba{2|4+6|5}\over\Spbb{2|1}\Spaa{6|5}P_{456}^2}-g^2\la
{\Spaa{1|5}\over\Spaa{6|5}P_{234}^2}+g^2\la{\Spbb{2|6}\over\Spbb{2|1}P_{345}^2},\nn
A_6(1^+,2,3,4,5,6^-)&=&g^4{\Spab{2|4+6|5}\over\Spaa{2|1}\Spbb{6|5}P_{456}^2}-g^2\la
{\Spbb{1|5}\over\Spbb{6|5}P_{234}^2}+g^2\la{\Spaa{2|6}\over\Spaa{2|1}P_{345}^2}.~~~\label{Yukawa-A6}
\eea
Before proceeding, several matters need to be emphasized.
Since a fermion propagator is ${\not P\over P^2}$, its choice of momentum direction
makes a difference, so we must insist one consistently.
Such a subtlety has been studied in \cite{Quigley:2005cu, Luo:2005rx, Georgiou:2004wu}.
Moreover, the following conventions are adopted:
Propagators of both fermions and scalars are without $i$'s.
For null momentum $-P$, we choose $\ket{-P}=\ket{P}$ and $\bket{-P}=-\bket{P}$.

\subsubsection{Four-point amplitude with two fermions}

The first example is amplitude $A_4(1^-,2,3,4^+)$. All possible physical
poles are $\{\Spbb{1|2}=\Spaa{3|4}\}$. From its expression in \eref{Yukawa-A4},
among $2C_4^2=12$ choices only $\Spab{4|1},\Spab{4|2},\Spab{3|1}$ are good deformations.

Let's start with bad deformation $\und{0}=\Spab{3|2}$, then the amplitude is
\bea
-A_4=g^2{\Spaa{1|4}\Spbb{1|4}\over\Spaa{2|4}\Spbb{2|1}}+{\cal B}^{\und{0}},~~~~\label{Yu-4p-1}
\eea
with the corresponding sets ${\cal D}^{\und{0}}=\{\Spbb{1|2}\}$,
${\cal U}^{\und{0}}=\emptyset$, ${\cal S}^{\und{0}}=\{\Spaa{2|4}\}$.

Although up to \eref{Yu-4p-1} all physical poles are detected, the spurious pole $\Spaa{2|4}$
has not been eliminated, thus we cannot set ${\cal B}^{\und{0}}=0$ yet.
Instead, let's perform deformation $\und{1}=\Spab{4|3}$ to detect $\Spaa{2|4}$.
Since its recursive part is zero, from \eref{Yu-4p-1} we find
\bea
{\cal B}{\cal R}^{\und{0},\und{1}}=-g^2{\Spaa{1|2}\Spaa{4|3}\Spbb{1|4}\over
\Spaa{2|3}\Spaa{2|4}\Spbb{2|1}}.~~~~~\label{Y-4p-c00}
\eea
It is worth mentioning that, although $\Spbb{2|1}$ and $\Spaa{2|3}$ appear in the denominator above,
the special combinations ${\Spbb{1|4}\over\Spaa{2|3}}$ and ${\Spaa{4|3}\over\Spbb{2|1}}$ prevent
them from being actual poles, which is again a safety check. Plugging \eref{Y-4p-c00} back, we get
\bea
-A_4=-g^2{\Spbb{2|4}\over\Spbb{2|1}}+{\cal B}^{\und{01}}.~~~~\label{Yu-4p-2}
\eea
Now above known term does not contain any spurious poles and all physical poles have been detected,
we can safely set ${\cal B}^{\und{01}}=0$, and reach the correct answer \eref{Yukawa-A4}.

\subsubsection{Five-point amplitude with two fermions}

The second example is amplitude $A_5(1^+,2,3,4,5^+)$.
Analysis of its factorization limits shows that all possible physical poles are
$\{\Spaa{1|2},\Spaa{4|5},\Spaa{5|1}\}$.
Here among $2C_5^2=20$ choices, only $\Spab{1|3},\Spab{1|4},\Spab{5|2},\Spab{5|3}$ are good deformations.

Let's start with bad deformation $\und{0}=\Spab{1|5}$, then the amplitude is
\bea
-A_5=g^3{\Spaa{2|4}\over\Spaa{2|1}\Spaa{5|4}}+{\cal B}^{\und{0}},~~~~\label{YU-5p-0}
\eea
with the corresponding sets ${\cal D}^{\und{0}}=\{\Spaa{1|2}\}$,
${\cal U}^{\und{0}}=\{\Spaa{4|5},\Spaa{5|1}\}$, ${\cal S}^{\und{0}}=\emptyset$.
Next, we perform deformation $\und{1}=\Spab{5|4}$, and its recursive part is
\bea
{\cal R}^{\und{1}}(z_{\und{1}})=g\la{1\over\Spaa{1|5}-z_{\und{1}}\Spaa{1|4}},
\eea
compare this with \eref{YU-5p-0}, the pole part gives
\bea
{\cal B}{\cal R}^{\und{0},\und{1}}=g\la{1\over\Spaa{1|5}}.
\eea
Plugging this back, we get
\bea
-A_5=g^3{\Spaa{2|4}\over\Spaa{2|1}\Spaa{5|4}}
-g\la{1\over\Spaa{5|1}}+{\cal B}^{\und{01}},~~~~\label{YU-5p-1}
\eea
with the corresponding sets
\bea
{\cal D}^{\und{01}}=\{\Spaa{1|2},\Spaa{5|1}\},~~~{\cal U}^{\und{01}}=\{\Spaa{4|5}\},~~~
{\cal S}^{\und{01}}=\emptyset.
\eea
To continue, we need to perform another deformation, \textit{e.g.}, $\und{2}=\Spab{5|1}$ to detect $\Spaa{4|5}$.
However, it can be checked that under $\und{2}$ the pole part of ${\cal B}^{\und{01}}$ is zero.
Since all physical poles have been detected, we can conclude that ${\cal B}^{\und{01}}=0$,
and the correct answer is
\bea
-A_5=g^3{\Spaa{2|4}\over\Spaa{2|1}\Spaa{5|4}}-g\la{1\over\Spaa{5|1}},~~~\label{YU-5p-2}
\eea
which matches \eref{Yukawa-A5}. Again only bad deformations have been used here.

\subsubsection{Six-point amplitude with two fermions}

The last example is amplitude $A_6(1^+,2,3,4,5,6^-)$.
Careful analysis of its factorization limits shows that all possible
physical poles are $\{\Spaa{1|2},\Spbb{5|6},P_{123}^2,P_{234}^2,P_{345}^2\}$.
And among $2C_6^2=30$ choices, the only good deformations are
$\Spab{1|i}$ with $i=3,4,5,6$, $\Spab{2|j}$ with $j=5,6$, and $\Spab{k|6}$ with $k=3,4$.

Let's start with bad deformation $\und{0}=\Spab{4|1}$, then the amplitude is
\bea
-A_6=-g^2\la{\Spaa{6|2}\over\Spaa{1|2}P_{345}^2}+g^2\la{\Spab{6|1+5|4}\over\Spab{1|5+6|4}P_{234}^2}
-g^4\left({\Spab{2|4+6|5}\over\Spaa{2|1}\Spbb{6|5}P_{123}^2}
+{\Spbb{4|5}\over\Spbb{6|5}\Spab{1|2+3|4}}\right)+{\cal B}^{\und{0}},~~~~\label{YU-6p-0}
\eea
with the corresponding sets
\bea
{\cal D}^{\und{0}}=\{P_{123}^2,P_{234}^2,P_{345}^2\},~~~
{\cal U}^{\und{0}}=\{\Spaa{1|2},\Spbb{5|6}\},~~~{\cal S}^{\und{0}}=\{\Spab{1|5+6|4},\Spab{1|2+3|4}\}.
\eea
Next, we perform deformation $\und{1}=\Spab{5|2}$ to detect
$\Spbb{5|6}$, $\Spab{1|5+6|4}$ and $\Spab{1|2+3|4}$, and its recursive part is
\bea
{\cal R}^{\und{1}}(z_{\und{1}})=-g^4{\Spab{2|4+6|5}\over\Spaa{2|1}\Spbb{6|5}P_{123}^2(z_{\und{1}})}
+g^2\la{\Spbb{1|5}\over\Spbb{6|5}P_{234}^2(z_{\und{1}})}
-g^2\la{\Spaa{6|2}\over\Spaa{1|2}P_{345}^2(z_{\und{1}})},
\eea
compare this with the pole part of \eref{YU-6p-0} (in fact, all known terms are pole parts), which gives
\bea
{\cal B}{\cal R}^{\und{0},\und{1}}=g^2\la{\Spbb{1|5}\over\Spbb{6|5}P_{234}^2}
-g^2\la{\Spab{6|1+5|4}\over\Spab{1|5+6|4}P_{234}^2}
+g^4{\Spbb{4|5}\over\Spbb{6|5}\Spab{1|2+3|4}}.
\eea
Thus we get
\bea
-A_6=-g^2\la{\Spaa{6|2}\over\Spaa{1|2}P_{345}^2}
+g^2\la{\Spbb{1|5}\over\Spbb{6|5}P_{234}^2}
-g^4{\Spab{2|4+6|5}\over\Spaa{2|1}\Spbb{6|5}P_{123}^2}+{\cal B}^{\und{01}},~~~~~~\label{YU-6p-1}
\eea
with the corresponding sets
\bea
{\cal D}^{\und{01}}=\{P_{123}^2,P_{234}^2,P_{345}^2,\Spbb{5|6}\},~~~
{\cal U}^{\und{01}}=\{\Spaa{1|2}\},~~~{\cal S}^{\und{01}}=\emptyset.
\eea
We can continue with deformation $\und{2}=\Spab{2|3}$ to detect the last
pole $\Spaa{1|2}$. It is easy to check that under $\und{2}$ the pole
part of ${\cal B}^{\und{01}}$ is zero. Since all physical poles have been
detected, we conclude that ${\cal B}^{\und{01}}=0$, and reach the correct answer by \eref{YU-6p-1}.

\section{Conclusion}

After presenting the new algorithm and several examples, we will give some remarks below.
Since this approach is based on pole structure of boundary contributions, and factorization limits
of amplitudes, which are general properties of quantum field theories, we expect that it
can be applied to theories with massive external and internal particles,
as well as theories in other dimensions. It will be interesting to explore these
directions. As mentioned before, if the boundary term does not contain any poles,
this approach can not be applied, which can happen for some effective theories.
More explicitly, if an effective theory has a primary interaction term as $\Phi^m$ in its Lagrangian,
an $m$-point amplitude will have a pure polynomial contribution for boundary terms. Then to determine
$n$-point amplitudes with $n>m$, we do need to know all these primary interaction vertices.

This algorithm is self-contained to calculate amplitudes without
beforehand analysis of the large $z$ behavior of an amplitude under a
given deformation, although knowing this will make calculations much easier.
In practical calculations, if the analysis of large $z$ behavior is
difficult, we can boldly use the on-shell recursion relation with
arbitrary choices of deformations. Then we need to judge whether the
result obtained is correct. There are three criteria:
(1) All spurious poles must be canceled out.
(2) The power of any physical pole must be at most one.
(3) It must have correct factorization limits for all physical poles.
If the result satisfies above all, it is very likely to be
correct (up to possible polynomial terms).
If it fails to satisfy at least one, there must
be missing boundary contributions, then one needs to continue the algorithm
to amend it until the correct answer is found.

At this point, we would like to discuss an important issue of the algorithm:
its efficiency. This paper concerns mostly with the feasibility of the
algorithm, as demonstrated by the examples. However, among all allowed
choices, how to choose deformations to maximally simplify the calculation
is still not clear. As shown by our cases,
although naively the algorithm requires to detect all possible
physical poles (maybe plus some spurious poles), in practice
quite often fewer steps than necessary can reach the complete answer.
Thus, how to optimize the choices is an important future project.

The idea behind this algorithm is quite simple and general, so it can be generalized
to many other cases. For example, so far we have only considered the BCFW deformation, but there
are other types such as the Risager deformation \cite{Risager:2005vk}.
It can also be applied to calculate the rational
parts of one-loop amplitudes \cite{Bern:2005hs, Bern:2005cq,
Berger:2006ci}. In this case double poles exist, for which the
general physical picture is not yet fully understood. Having
determined the full tree amplitudes, and after a combined use of
unitarity cut \cite{Bern:1994zx,Britto:2004nc}\footnote{When combining tree amplitudes
together along cuts, there are a few ambiguities to be fixed. The similar idea was
used in \cite{Zhou:2014yaa} recently, where tree amplitudes are reconstructed
by combining all factorization channels.}, it is
possible to reconstruct loop amplitudes without recourse to Feynman
diagrams. Furthermore, it is also intriguing to apply it
to generalize the recursion relation for loop integrands of ${\cal N}=4$
SYM \cite{ArkaniHamed:2010kv}, to general quantum field theories.

\section*{Acknowledgement}

The authors would like to thank David Kosower and Qingjun Jin for valuable
discussions. This work is supported, in part,
by Qiu-Shi\footnote{Qiu-Shi in Chinese means `to explore the truth'.} funding and
Chinese NSF funding under contracts No.11031005, No.11135006 and No.11125523.

\appendix
\section{More Abstract Operation}

In this appendix, we introduce a concise algebraic language to reinterpret the
new algorithm. Assume that $A$ is a rational function
of external kinematic variables $\la_i$,$\W\la_i$,
its poles can be {\sl physical or spurious} and we treat them on the same footing.
Consider its deformed version $A(z_{\underline{s}})$ with the deformed spinor pair
$\Spab{i_s|j_s}$, as a rational function of $z_{\underline{s}}$,
$A(z_{\underline{s}})$ is decomposed in the same way as \eref{Az0-gen}, namely
\bea
A(z_{\underline{s}})=\sum_{P^2_t\in{\cal D}^{\underline{s}}}
{a_t\over P^2_t(z_{\underline{s}})}+C^{\underline{s}}_0
+\sum C^{\underline{s}}_iz_{\underline{s}}^i,~~~~\label{Az-split}
\eea
where again ${\cal D}^{\underline{s}}$ and ${\cal U}^{\underline{s}}$
denote the detectable and undetectable poles of deformation $\underline{s}$ respectively.
After the decomposition, one can define two operators as below
\bea
{\cal P}^{\underline{s}}[A]\equiv
\sum_{P^2_t\in{\cal D}^{\underline{s}}}{a_t\over P^2_t(z_{\underline{s}}=0)},~~~~~
{\cal C}^{\underline{s}}[A]\equiv C_0.~~~\label{def-PC}
\eea
By this definition the identity operator can be trivially written as
\bea
{\cal I}={\cal P}^{\underline{s}}+{\cal C}^{\underline{s}}.~~~\label{iden}
\eea
Now the new algorithm can be rewritten as the following:
Starting from \eref{iden}, one can insert the identity operator into any place as pleased. One way is
\bea
{\cal I}&=&{\cal P}^{\underline{0}}+{\cal C}^{\underline{0}}
={\cal P}^{\underline{0}}+{\cal I}{\cal C}^{\underline{0}}
={\cal P}^{\underline{0}}+({\cal P}^{\underline{1}}
+{\cal C}^{\underline{1}}){\cal C}^{\underline{0}}\nn
&=&{\cal P}^{\underline{0}}+{\cal P}^{\underline{1}}{\cal C}^{\underline{0}}
+{\cal C}^{\underline{1}}{\cal C}^{\underline{0}},~~~~\label{use1}
\eea
which exactly matches \eref{Az0-gen-1}, where ${\cal B}{\cal R}^{\und{0},\und{1}}
={\cal P}^{\underline{1}}{\cal C}^{\underline{0}}[A]$ and
${\cal B}^{\und{01}}={\cal C}^{\underline{1}}{\cal C}^{\underline{0}}[A]$.
In the other way, \eref{Az0-gen-2} can be recovered by
\bea
{\cal I}&=&{\cal P}^{\underline{1}}+{\cal C}^{\underline{1}}
={\cal P}^{\underline{1}}+{\cal C}^{\underline{1}}{\cal I}
={\cal P}^{\underline{1}}+{\cal C}^{\underline{1}}({\cal P}^{\underline{0}}
+{\cal C}^{\underline{0}})\nn
&=&{\cal P}^{\underline{1}}+{\cal C}^{\underline{1}}{\cal P}^{\underline{0}}
+{\cal C}^{\underline{1}}{\cal C}^{\underline{0}},~~~~\label{use2}
\eea
where ${\cal R}{\cal B}^{\und{0},\und{1}}={\cal C}^{\underline{1}}{\cal P}^{\underline{0}}[A]$
and ${\cal B}^{\und{01}}={\cal C}^{\underline{1}}{\cal C}^{\underline{0}}[A]$.
The pattern of the new algorithm is then clear: After inserting identity
operators repeatedly into terms with ${\cal C}$'s only, one can
expand them to specify the calculation at each step. For the
${\cal P}$'s at rightmost, we know how to calculate by using the
recursion relation. For the ${\cal C}$'s following after ${\cal P}$'s,
the constant extraction is also known. And for the last term with ${\cal C}$'s only,
namely ${\cal C}^{\underline{n}}{\cal C}^{\underline{n-1}}\ldots
{\cal C}^{\underline{1}}{\cal C}^{\underline{0}}$,
when the $n$-th step covers all possible physical poles and there is no spurious pole
in all known terms, one can finally set it to zero (up to possible polynomial terms).


\end{document}